\documentclass[12pt,preprint]{aastex}
\usepackage{emulateapj5}
\usepackage{apjfonts}

\shorttitle{Non-detection of H$_3^+$ in HD~141569A}
\shortauthors{Goto et al.} 

\begin{document}

\title{Search for H$_3^+$ in HD~141569A\altaffilmark{1,2}}

\author{Miwa Goto,\altaffilmark{3}
  T. R. Geballe,\altaffilmark{4}
  B. J. McCall,\altaffilmark{5}
  T. Usuda,\altaffilmark{6} \\
  H. Suto,\altaffilmark{7} 
  H. Terada,\altaffilmark{6} 
  N. Kobayashi,\altaffilmark{8} \\ 
\and T. Oka\altaffilmark{9}}

\email{mgoto@mpia-hd.mpg.de}

\altaffiltext{1}{Based on data collected at the Subaru Telescope, which is
                 operated by the National Astronomical Observatory of
                 Japan.}

\altaffiltext{2}{Based on data collected at the United Kingdom Infrared 
		Telescope, which is operated by the Joint Astronomy Centre
		on behalf of the U.K. Particle Physics and Astronomy
		Research Council.}
     
\altaffiltext{3}{Max Planck Institute for Astronomy, Koenigstuhl 17,
                 D-69117 Heidelberg, Germany.}

\altaffiltext{4}{Gemini Observatory,
                 670 North A`ohoku Place,  Hilo, HI 96720.}

\altaffiltext{5}{Departments of Chemistry and Astronomy,
                 University of Illinois at Urbana-Champaign,
                 Urbana, IL 61801-3792.}

\altaffiltext{6}{Subaru Telescope, 650, North A`ohoku Place, Hilo, HI
  96720.}

\altaffiltext{7}{National Astronomical Observatory of Japan, Osawa,
Mitaka, Tokyo 181-8588, Japan}

\altaffiltext{8}{The University of Tokyo, Osawa, Mitaka,
Tokyo 181-0015, Japan}

\altaffiltext{9}{Department of Astronomy and Astrophysics,
                 Department of Chemistry, and Enrico Fermi Institute,
                 University of Chicago, Chicago, IL 60637.}

\begin{abstract}

A search for H$_3^+$ line emission, reported to have been detected
toward the young star HD~141569A and possibly originating in a clump
of planet-forming gas orbiting the star, has yielded negative results.
Observations made at the United Kingdom Infrared Telescope and at the
Subaru Telescope during 2001-2005 covered 11 major transitions of
H$_3^+$ from 3.42 to 3.99~$\mu$m. No H$_3^+$ emission lines were
detected; one marginal detection at 3.9855~$\mu$m in June 2002 was not
confirmed in later spectra. The upper limits to the line strengths are
significantly lower than the previously reported
detections. Supplemental slit-scanning spectroscopy using adaptive
optics was performed within 0\farcs38 of HD~141569A to search for
extended emission from H$_3^+$, but no emission was detected. We
compare our upper limit to the luminosity in H$_3^+$ from HD~141569A
with that possible from a gas giant protoplanet and also from a jovian
mass exoplanet in close orbit about its central star.

\end{abstract}


\section{Introduction}

\figurenum{1}
\begin{figure*}[tbh]
\begin{center}
\includegraphics[angle=-90,scale=0.75]{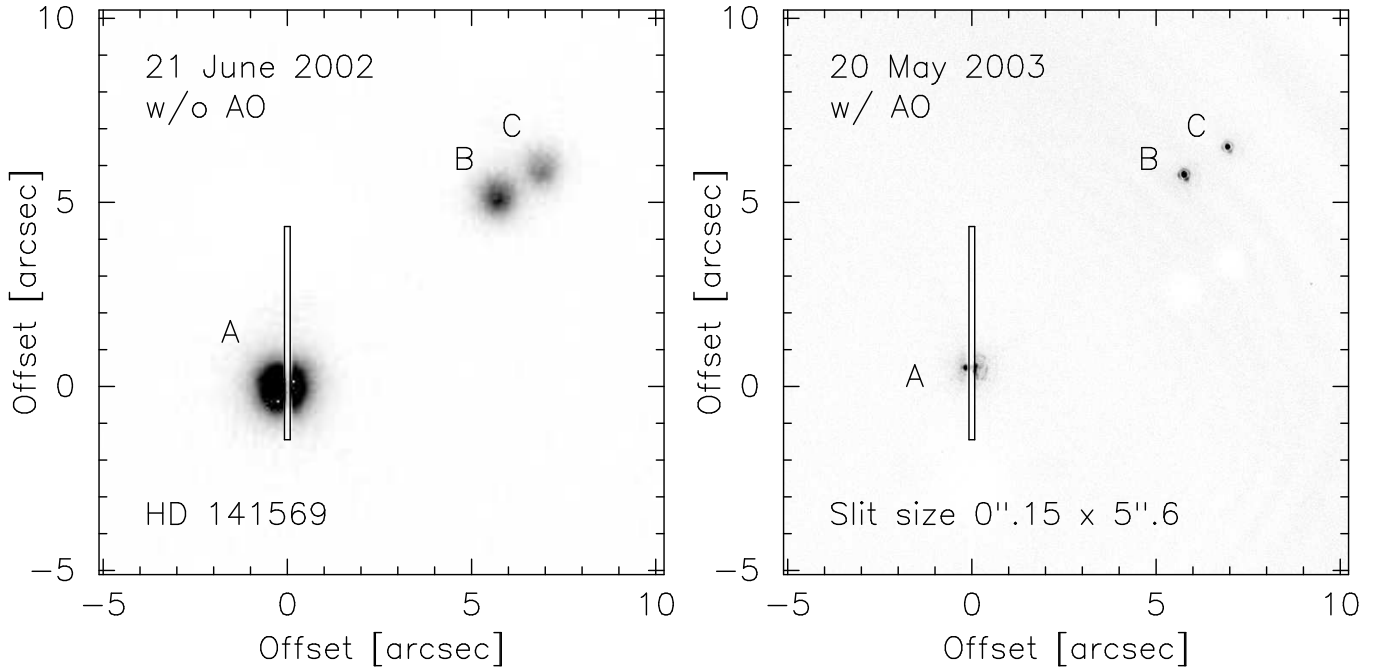}
\caption{Slit positions at Subaru overlaid with the images obtained by the
slit viewing camera. HD~141569 is a triple system consisting of primary
HD~141569A with faint companions B and C at 8\farcs6 northwest
(Weinberger et al. 2000). Left: $K$-band image on 21 June 2002 under natural
seeing conditions.  Right: $L'$-band image on 20 May 2003 obtained with 
the adaptive optics system. Offsets were performed as shown to search for
for H$_3^+$ line emission detached from the central star. 
\label{f1}}
\end{center}
\end{figure*}

The detection of the $Q$(1,~0) and $Q$(3,~0) emission lines of
vibrationally excited H$_3^+$ near 4.0~$\mu$m was recently reported by
Brittain \& Rettig (2002; hereafter BR02) toward the nearby star
HD~141569A. Apart from a highly tentative detection of H$_3^+$ in
SN1987A (Miller et al. 1992), this is the first claim of H$_3^+$ {\it
in emission} from beyond the solar system. HD~141569A is a young star
with a debris disk that has been extensively observed in the visible
to mid-infrared region (Augereau et al. 1999; Weinberger et al. 1999;
Fisher et al 2000; Mouillet et al. 2001; Marsh et al. 2002, Brittain
et al. 2003; Clampin et al. 2003). The \ion{H}{1} lines in emission in
the visible part of the spectrum indicate high chromospheric activity
(Andrillat, Jaschek, \& Jaschek 1990). The small infrared excess
($L_{\rm IR}/L_{\ast} = 8\times 10^{-3}$; Sylvester \& Skinner 1996)
of the object is comparable to other debris disk objects ($L_{\rm
IR}/L_{\ast} = $ $10^{-3}$--$10^{-4}$), implying that the dust in
HD~141569A is dissipating and already mostly removed (Malfait,
Bogaert, \& Waelkens 1998). The low abundance of circumstellar CO
(Zuckerman, Forveille, \& Kastner 1995) also suggests that HD~141569A
is approaching the zero-age main sequence and is currently in
transition from a Herbig Ae/Be star to a gas-exhausted Vega-type
object. Weinberger et al. (2000) used spectroscopy of the low-mass
companions of HD~141569A to constrain the age of the system to
5$\pm$3~Myr, between those of Herbig Ae/Be stars and the archetypical
Vega-type star $\beta$~Pic ($\sim$20~Myr; Barrado y Navascu\'{e}s et
al. 1999). The time interval of 1--10~Myr after star formation is
critical for planet formation since the building of Jupiter-like
planets is believed to take place during that period (e.g. Lissauer
1993; Pollack et al. 1996).  Observations of HD~141569A and other
transition objects are needed to understand if, when, and how gas
giant planets form around the intermediate mass stars.

\tablenum{1}
\begin{table*}[b]
\begin{center}
\caption{Observing log. \label{tb1}}
\begin{tabular}{llccccrcll}
\tableline\tableline

    {UT Date               } &
    {Line                  } &
    {Telescope             } &
    {Instrument            } &
    {Resolution            } &
\multicolumn{2}{c}{Slit        } &

    {Integration\tablenotemark{a}} &
    {Atmospheric Std.     }  \\

    {                      } &
    {                      } &
    {                      } &
    {                      } &
    {                      } &
    {Width                 } &
    {P.A.                  } &
    {                      } &
    {                      }\\


\tableline

2001 Sep 30 & $R$(1,~1)$^u$, $R$(1,~0)        & UKIRT  & CGS4 & $R=$33,000 & 0\farcs45 & 90\arcdeg & 24 min.  & HR~5892 (A2Vm)\\
2002 Jun 21 & multiple lines\tablenotemark{b} & Subaru & IRCS & $R=$20,000 & 0\farcs15 &  0\arcdeg & 96 min.  & HR~6194 (A3IV)\\
2002 Jun 21 & multiple lines                  & Subaru & IRCS & $R=$20,000 & 0\farcs15 & 90\arcdeg & 96 min.  & HR~6194       \\
2003 May 20 & multiple lines                  & Subaru & IRCS+AO & $R=$20,000 & 0\farcs15 &  0\arcdeg & 24 min.  & HR~5685 (B8V) \\
2003 May 20 & multiple lines                  & Subaru & IRCS+AO & $R=$20,000 & 0\farcs15 &  0\arcdeg & 60 min.\tablenotemark{c} & HR~5685 \\
2005 Mar 2-4 & $Q$(3, 0)                      & UKIRT  & CGS4 & $R=$21,000  & 0\farcs8  & 90\arcdeg & 285 min & HR~5511 (A0V) \\

\tableline
\end{tabular}

\tablenotetext{a}{On-source integration time.}  
\tablenotetext{b}{Includes $R$(3,~3)$^u$, $R$(4, 4)$^l$, $R$(1,~1)$^u$, 
$R$(1,~0), $R$(1,~1)$^l$, $Q$(2,2), $Q$(1,~1), $Q$(1,~0), $Q$(2,~1)$^l$, 
$Q$(3,~0) and $Q$(3,~1)$^l$.}
\tablenotetext{c}{Sum of slit scans across $\pm$0\farcs38 at
  HD~141569A except for the exposures at the position of the star.}

\end{center}
\end{table*}

Although interstellar H$_3^+$ has been detected in absorption for
almost a decade (Geballe \& Oka 1996; McCall et al. 1998, 1999, 2002;
Geballe et al.  1999, Goto et al. 2002, Brittain et al. 2004), the
molecular ion has never been definitively found in emission except in
the planetary atmospheres of our solar system.  The most prominent
H$_3^+$ emission in the solar system is the aurora at the Jovian poles
(Drossart et al. 1989; Oka \& Geballe 1990). Miller et al. (2000)
discussed the feasibility of detecting H$_3^+$ in giant exoplanets
brought into close orbits around their stars, and concluded that
detection would be difficult.

The H$_3^+$ line intensities reported by BR02 are remarkably high
compared to those of Jupiter. From the equivalent width of the
$Q$(3,~0) line, the peak of which BR02 observed to be roughly 5\%
above the continuum of HD~141569A, and the distance of 99~pc to
HD~141569A (van den Ancker et al. 1998), the total luminosity of that
line alone is $\sim$3~$\times$~10$^{19}$~W. In comparison, the total
H$_3^+$ luminosity from Jupiter (Miller et al. 2000) is $\sim$10$^{7}$
times less. This huge disparity demonstrates that the putative line
emission from HD~141569A cannot arise in a giant planet. BR02
suggested that the line-emitting H$_3^+$ may be distributed in a huge
gas giant protoplanet with a volume of $\sim$1 AU$^{3}$ or at the
inner edge of the circumstellar disk at a distance no less than 17 AU
from the star, but did not identify physical mechanisms that might
lead to the production of the required amount of vibrationally excited
H$_3^+$.

The goals of the observations reported here were to confirm the presence
of H$_3^+$ in HD~141569A and obtain better diagnostics of the physical
state of the circumstellar gas of HD~141569A by examining a wide range of
H$_3^+$ and possibly other transitions in the 3--4~$\mu$m region. In the
following sections we describe our observations and data reduction, report
the results, and compare them with H$_3^+$ line intensities that might be
expected in a variety of situations.

\section{Observations}

Initial observations of HD~141569A were made at the United Kingdom
Infrared Telescope (UKIRT) on 2001 September 30 using the facility
1--5~$\mu$m spectrometer CGS4 (Mountain et al. 1990). More
comprehensive searches were made at the Subaru Telescope on 2002 June
21 and 2003 May 20 with the facility's Infrared Camera and
Spectrograph (IRCS; Tokunaga et al. 1998; Kobayashi et al. 2000). An
additional measurement using UKIRT was made in 2005 March. An
observing log is provided in Table~\ref{tb1}.

\figurenum{2}
\begin{figure*}[tbh]
\begin{center}
\includegraphics[angle=-90,scale=0.75]{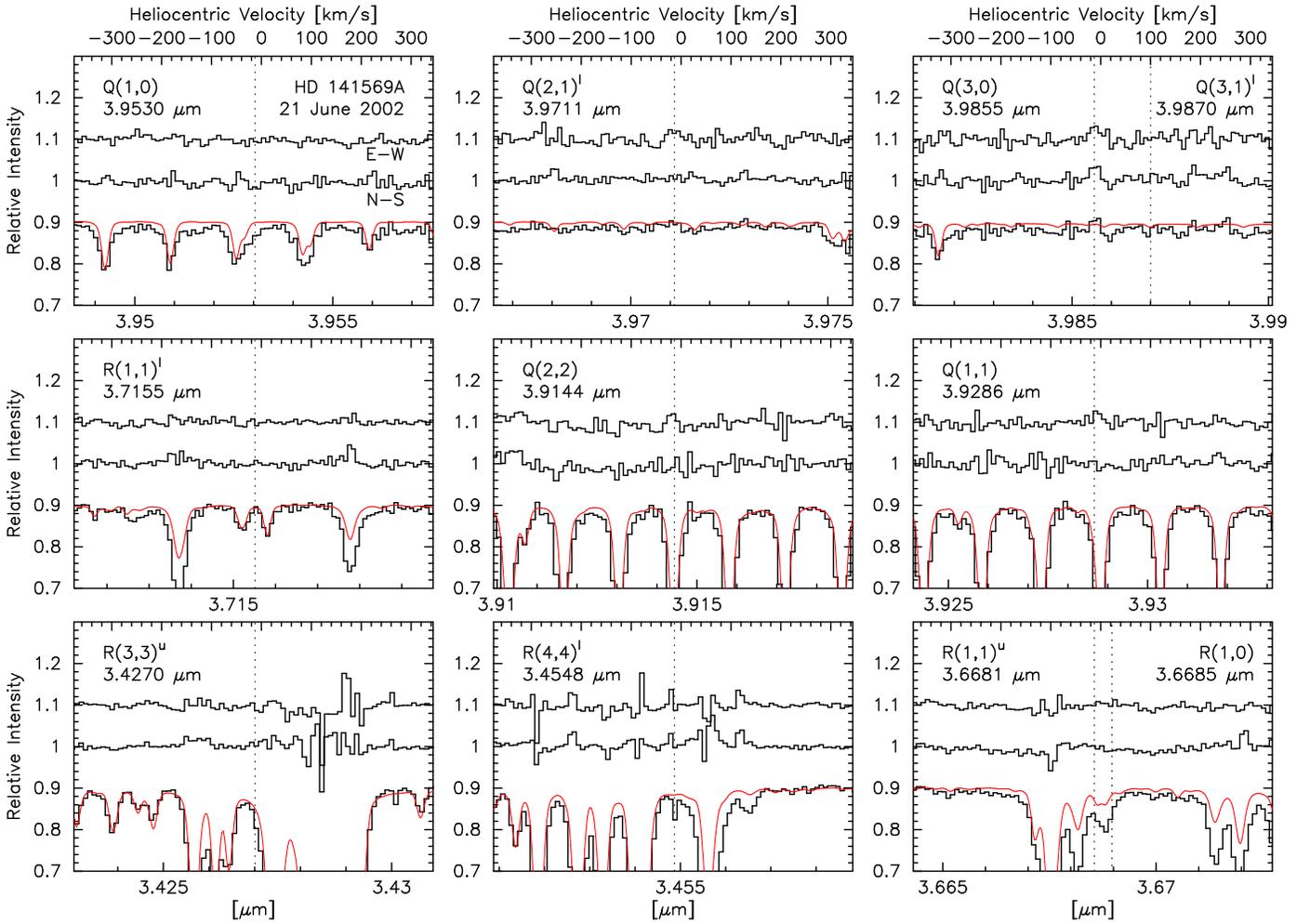}
\caption{Subaru spectra of HD~141569A on 21 June 2002. Spectral
segments at each H$_3^+$ transition are shown in individual
panels. The upper spectrum in each panel was recorded with the slit
aligned east-west, the middle spectrum is with the slit north-south
(Fig.~\ref{f1}). The bottom trace in black is the north-south spectrum
before ratioing with the standard star spectrum. The red overlay is
the atmospheric transmission curve calculated by ATRAN (Lord 1992).
Spectra are vertically offset by $\pm$0.1.  Dotted lines are at
expected positions of H$_3^+$ lines (Lindsay \& McCall 2001) and
include corrections for a heliocentric velocity of $-$13~km~s$^{-1}$
(BR02) and the earth's orbital motion.
\label{f2}}
\end{center}
\end{figure*}

\figurenum{3}
\begin{figure*}[tbh]
\begin{center}
\includegraphics[angle=-90,scale=0.75]{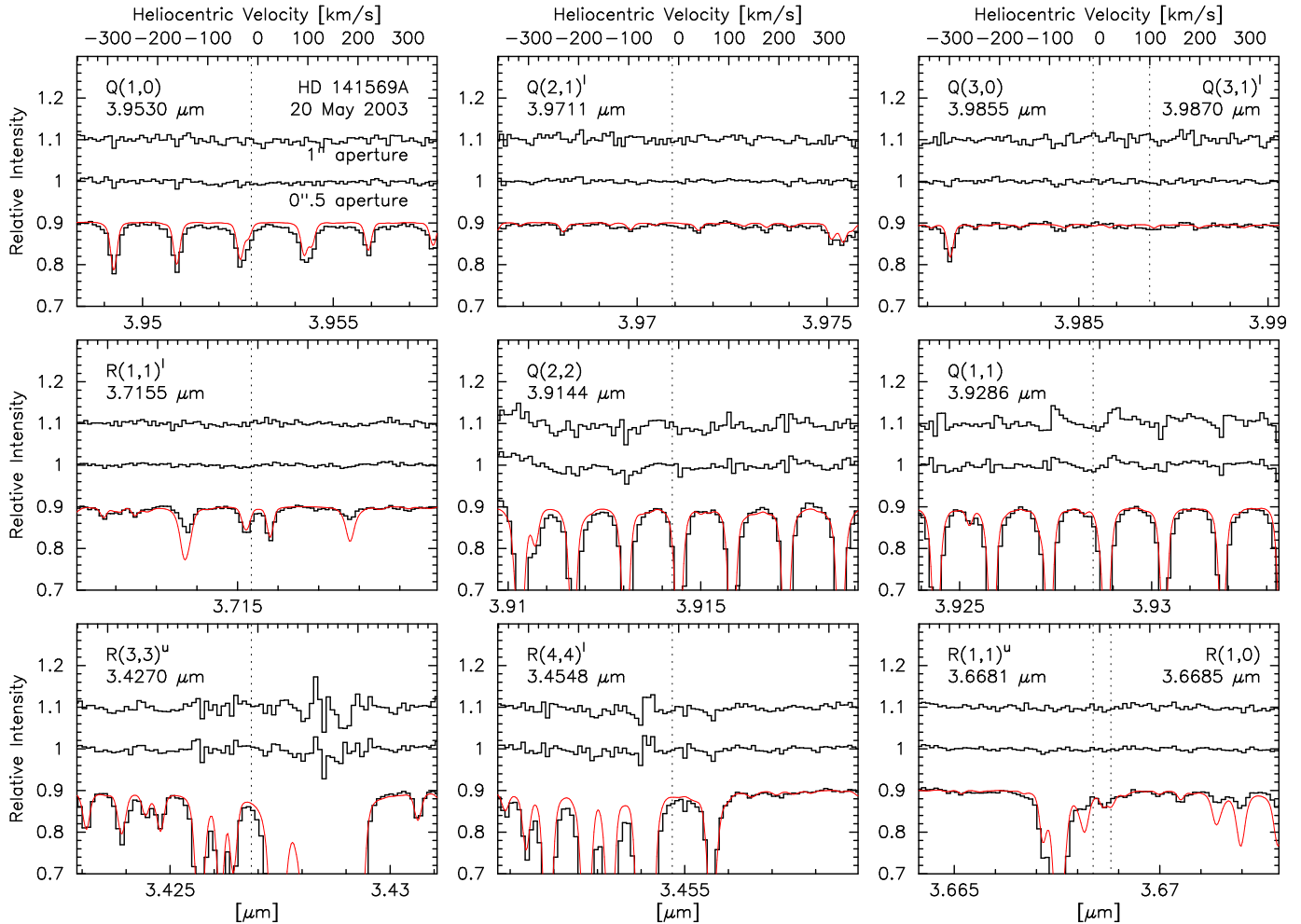}
\caption{Same as Fig. \ref{f2}, but for the second run on 20 May 2003
with an adaptive optics system. Upper trace is extracted from a 1\farcs0
aperture; middle trace is from 0\farcs5 aperture. The bottom trace in
black is the spectrum with 0\farcs5 aperture before ratioing with the
standard star spectrum.\label{f3}}
\end{center}
\end{figure*}

\figurenum{4}
\begin{figure*}[tbh]
\begin{center}
\includegraphics[angle=-90,scale=0.75]{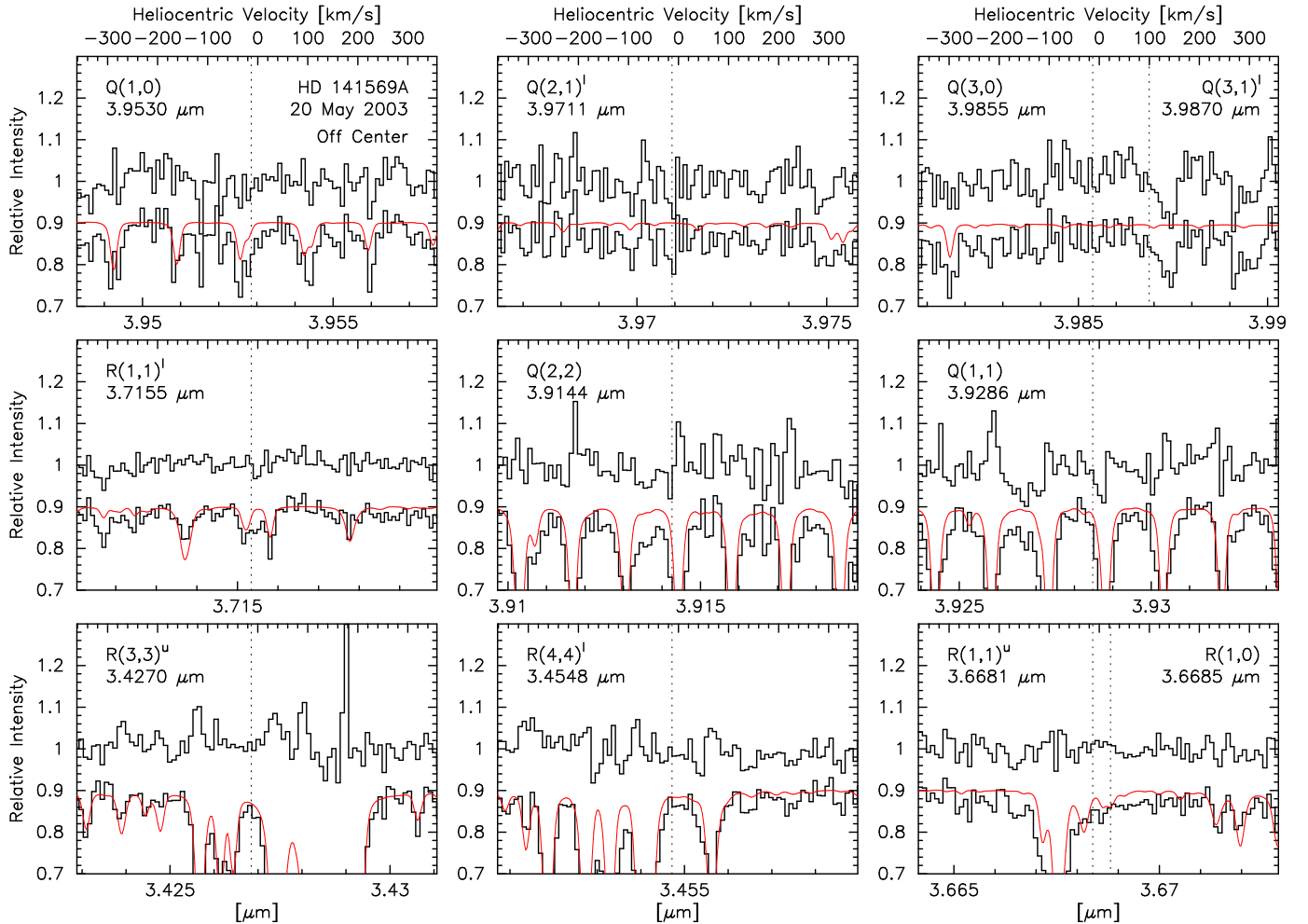}
\caption{Same as Fig. \ref{f2}, but for the slit-scanning
spectroscopy conducted during the second run. The signals in the
off-center region $\pm$0\farcs38 in the east and the west of
HD~141569A are summed in the upper traces; the lower traces are the
unratioed spectra. The continuum is due to spill-over from the central
star and is roughly a quarter of that in Fig. \ref{f3}.
\label{f4}}
\end{center}
\end{figure*}

The UKIRT observations in 2001 utilized CGS4's echelle to cover the
region of the H$_3^+$ $R$(1,~1)$^{u}$ $R$(1,~0) ortho-para doublet at
3.67~$\mu$m, with the 0\farcs45 wide slit of the spectrograph oriented
east-west. The resolving power of the instrument in this mode was 33,000.
The measurements and the data reduction were performed in a manner similar
to those reported by Geballe \& Oka (1996) and Geballe et al. (1999). The
spectrum of HR~5892, observed prior to and after HD~141569A, was used to
remove telluric absorption lines and provided wavelength calibration
accurate to 8~$\times$~10$^{-6}$~$\mu$m rms. 

At Subaru the echelle and a cross-dispersing grating in the IRCS were
employed to cover one-third of the wavelength range from 3.22 to
4.01~$\mu$m. The grating angles were set to record the following major
H$_3^+$ transitions in one exposure: $R$(3,~3)$^u$, $R$(4,~4)$^l$,
$R$(1,~1)$^u$, $R$(1,~0), $R$(1,~1)$^l$, $Q$(2,~2), $Q$(1,~1),
$Q$(1,~0), $Q$(2,~1)$^l$, $Q$(3,~0) and $Q$(3,~1)$^l$, which includes
the two lines reported by BR02. The narrow slit (0\farcs15 $\times$
5\farcs6) was used to obtain spectra with a resolving power of
$R$=20,000. The slit position angle was set to 0\arcdeg~and 90\arcdeg~
in the first run in June 2002 (Fig. \ref{f1}). The telescope was
nodded 2\farcs8~along the slit to allow subtraction of the sky
emission and the dark current. The on-source integration time for each
slit angle was 96 min. An early-type star HR~6149 (A0V) was observed
immediately after HD~141569A as a spectroscopic standard. A
spectroscopic flat field was obtained at the end of the night using a
halogen lamp which is in the calibration unit at the telescope's
cassegrain port in front of the entrance of the spectrograph.

The second Subaru observing run was carried out in the same manner,
but with the facility adaptive optics (AO) system (Gaessler et al.
2002; Takami et al. 2004), a 36-element curvature-base system that
delivers nearly diffraction-limited images at wavelengths longer than
2~$\mu$m. The AO system improved the transmission through the slit by
2--4 times over the seeing-limited condition. The slit position angle
was aligned north-to-south, along the apparent major axis of the
circumstellar disk. A two-point dithering sequence was provided by a
tip-tilt mirror inside the AO system. The on-source integration time
was 24~min. Supplemental slit-scanning spectroscopy was performed over
an area 0\farcs38 on each side of HD~141569A which is $\pm$38~AU in
projection at the distance of the object. The telescope was moved to
blank sky 10\arcsec~north of HD~141569A in between the exposures at
each slit position. HR~5685 (B8V) was observed as an atmospheric
standard star at the similar airmass with the program star. The
spectroscopic flat field was obtained during the night immediately
after HD~141569A, without moving the telescope or the instrument
rotator, in order to minimize the fringes in the spectrum. 

CGS4 was used at UKIRT during 2005 March 2-4 to search for the
$Q$(3,~0) line of H$_3^+$ at 3.985~$\mu$m, with HR~5511 as the
telluric standard.  Because of poor seeing conditions a somewhat wider
slit (0\farcs85) was employed than in 2001, producing a resolving
power of 21,000.

\section{Reduction of Subaru Data}

Consecutive frames recorded in the dithering sequence at Subaru were
subtracted to remove the sky emission, and the subtracted frames were
stacked together. For slit-scanning spectroscopy blank sky
echellograms were subtracted. The pixel-by-pixel variation of the
detector response was normalized by ratioing with dark-subtracted flat
field images. Outlier pixels were determined based on the statistics
of the dark current and flat field images. The signals in these pixels
were determined by interpolation before extracting one dimensional
spectra. All the above procedures were handled with IRAF\footnote{IRAF
is distributed by the National Optical Astronomy Observatories, which
are operated by the Association of Universities for Research in
Astronomy, Inc., under cooperative agreement with the National Science
Foundation.} packages for image reduction and aperture extraction.

Wavelength calibration was performed by maximizing the
cross-correlation of the observed spectra with the atmospheric
transmission curve modeled by ATRAN (Lord 1992). The uncertainty of
the wavelength calibration is dependent on the local line density, but
typically is better than a tenth of a pixel $<$1$\times 10^{-5}~\mu$m
(0.75~km~s$^{-1}$). Further processing included linear registration of
standard star and the object spectra, rescaling of the standard star
spectrum according to Beer's law to minimize the mismatch of the
airmass, convolving the spectrum to equalize the slightly different
spectral resolutions, and finally dividing by the standard star
spectrum to eliminate the atmospheric absorption lines. Some of the
fully processed data showed periodic wavy features with amplitude
smaller than 2\% of the continuum flux. This fringing pattern was
removed by discarding a few protruding frequencies in the
Fourier-transformed spectra.

The reduced Subaru spectra from June 2002 and May 2003 are shown in
Fig. \ref{f2} and Fig. \ref{f3}, respectively. Both the north-south
and east-west slit spectra are shown in Fig. \ref{f2}; in
Fig. \ref{f3} spectra extracted from 0\farcs5 and 1\farcs0 apertures
along the slit are shown. The slit-scanning spectra obtained in the
second run are shown in Fig. \ref{f4}. The data from off-center
regions to the east and the west of HD~141569A were summed in order to
provide a higher signal-to-noise ratio.

\section{Results}

No emission lines were detected at the wavelengths of any of the
H$_3^+$ transitions covered by these observations. In fact, no emission
lines whatsoever were detected at any of the wavelengths covered. Limits
for unresolved and maginally resolved lines are typically a few percent of
the continuum.

\figurenum{5}
\begin{figure*}[tbh]
\begin{minipage}[b]{8.8cm}
\begin{center}
\includegraphics[angle=0,scale=0.4]{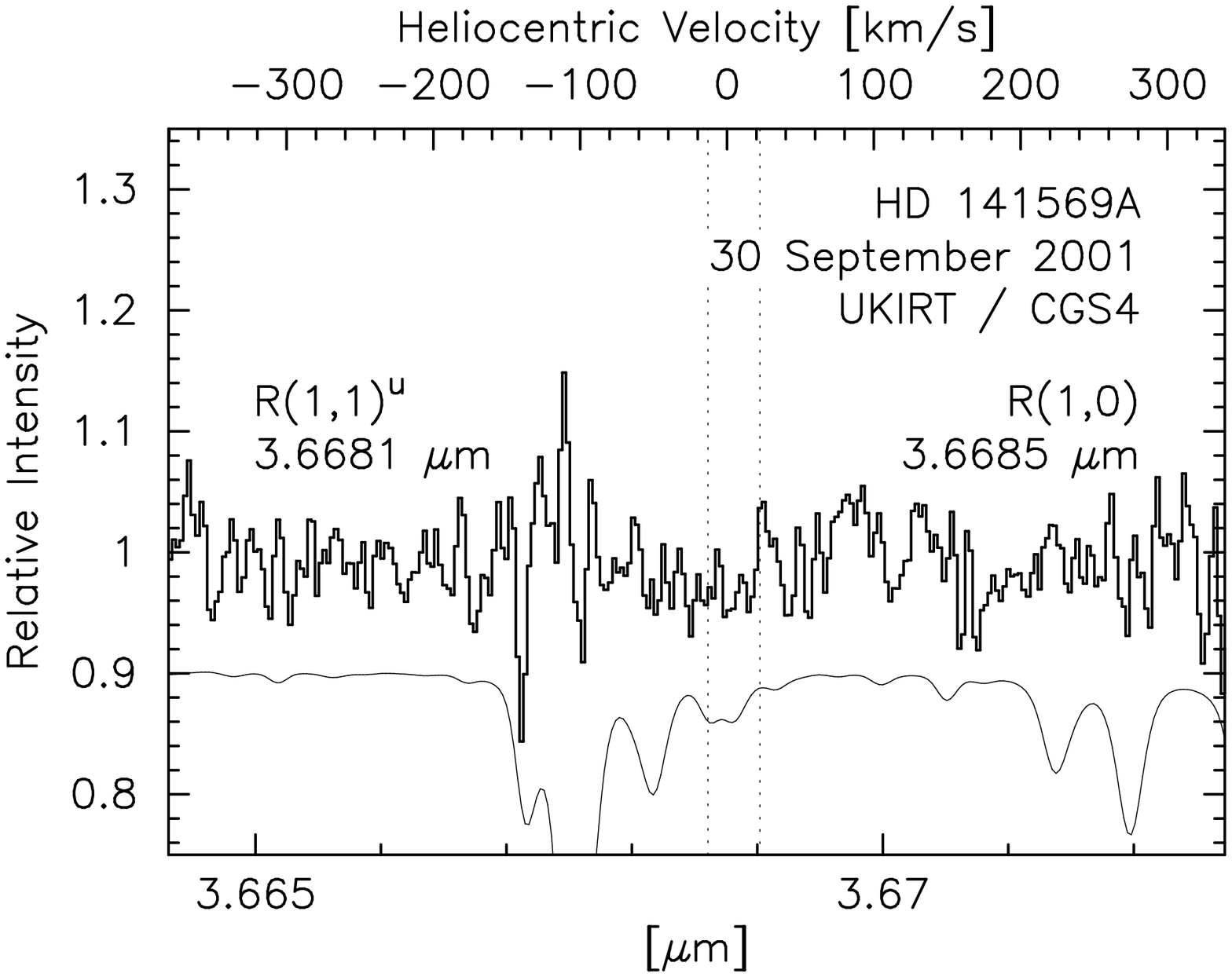}
\caption{$R$=33,000 spectrum of HD~141569A at the $R$(1,~1)$^u$ and
$R$(1,~0) doublet observed on 30 September 2001 at UKIRT. The
expected line positions of the doublet at the time of observing are
marked with dotted vertical lines. 
\label{f5}}
\end{center}
\end{minipage}
\figurenum{6}
\hspace{0.5cm}
\begin{minipage}[b]{8.8cm}
\begin{center}
\includegraphics[angle=0,scale=0.5]{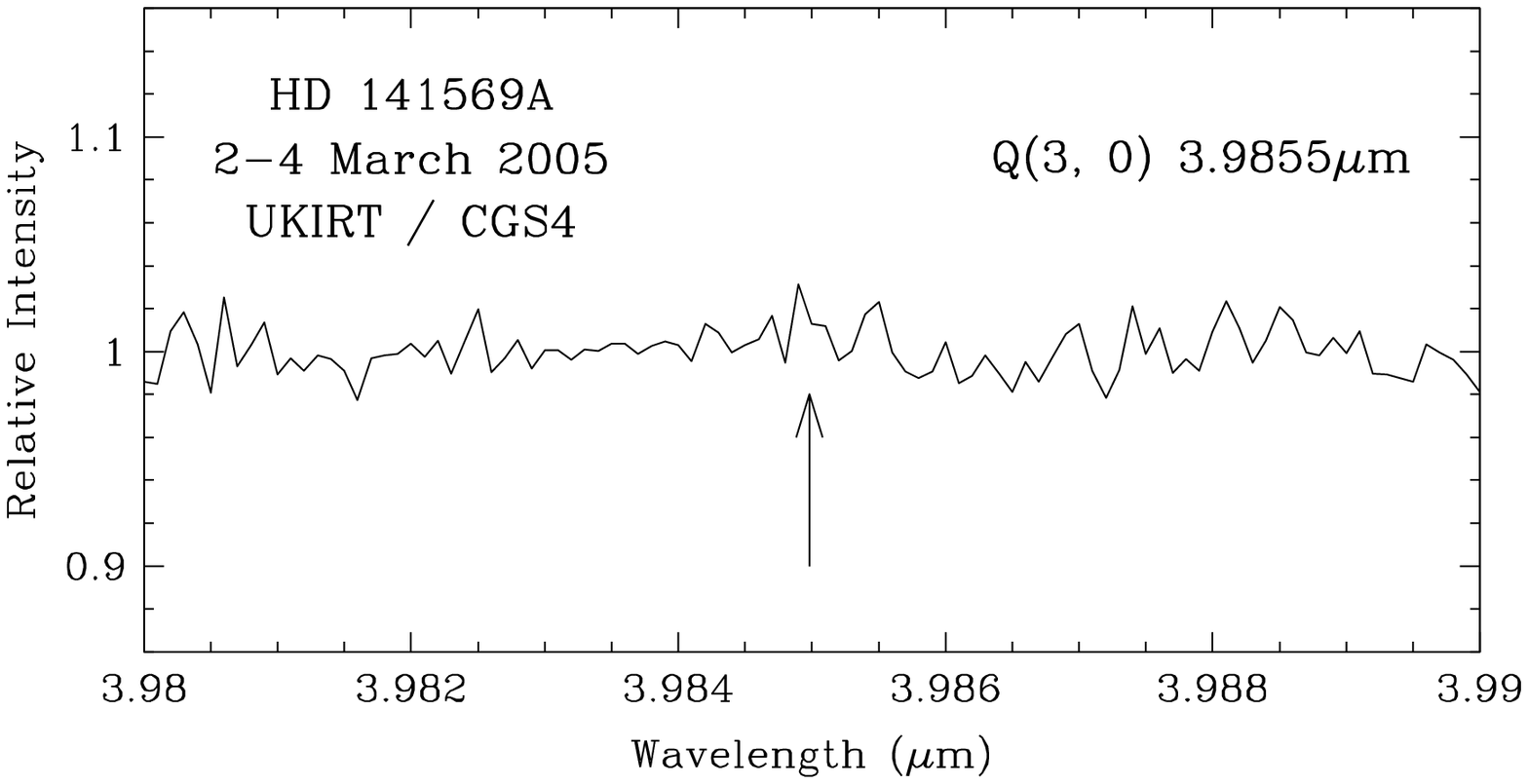}
\caption{$R$=21,000 spectrum of HD~141569A at the $Q$(3,~0) line observed
on 1-3 March 2005 at UKIRT. The expected line position is marked with an
arrow.\label{f6}}
\end{center}
\end{minipage}
\end{figure*}

The sensitivity of each of the Subaru observations was estimated using
the standard deviation of the fluctuations in the continuum. The
statistics were obtained over the data points shown in each panel in
Fig. \ref{f2} and \ref{f3}. The detection limits per resolution
element (2~pixels) for the emission lines are 4--5\% (3$\sigma$) for
the first run in June 2002, and $\sim$2\% for the second run in May
2003, except for those few transitions having severe overlap with
strong telluric absorption lines.  The 3$\sigma$ noise levels in the
UKIRT spectra are $\sim$4\% of the continuum per resolution element (3
pixels) near the ortho-para doublet (except on the strong telluric
CH$_{4}$ complex at 3.6675~$\mu$m) and $\sim$0.8\%. of the continuum
per resolution element (2 pixels) near the $Q$(3,~0) line.

The two $Q$-branch lines in the spectrum reported by BR02 are strong
enough that they would have been seen clearly in the Subaru and UKIRT
spectra, which were obtained at similar resolutions as those obtained
by BR02. In addition to this, the absences of many of the other
emission lines, in particular the $R$(1, 1)$^{u}$ $R$(1,~0) ortho-para
doublet at 3.67~$\mu$m, which was not detected either at Subaru or at
UKIRT (Fig.  \ref{f5}, upper panel), and also not by BR02, are
difficult to reconcile with the reported detections of the $Q$(1,~0)
and $Q$(3,~0) lines. In Jupiter all of those lines are observed with
comparable intensities (Maillard et al. 1990). The $Q$(1,~0) and
$Q$(3,~0) lines originate in ($J$, $G$) = (1,~0) and (3,~0) levels of
the vibrationally excited state, respectively. The accidental doublet,
consisting of the $R$(1,~0) and R(1,~1)$^{u}$ transitions, arises from
the ($J$, $G$) = (2,~0) and (2,~1) levels in the $\nu$=1 state. All
known plasma sources of H$_3^+$ produce fairly thermal distributions
of population, with the exception of an enhancement of metastable
levels in very low density media (Oka \& Epp 2004). We are not aware
of a physical mechanism that could explain the emission from the
vibrationally excited (1,~0) and/or (3,~0) states without causing
emission from the excited (2,~0) state.

There is a modest bump near the wavelength of the $Q$(3,~0) line in
the June 2002 Subaru spectrum, which is marginally above the noise
level and roughly half the strength of the line reported by BR02. A
possible identification of the bump could be \ion{C}{1} at
3.98546~$\mu$m (2509.12~cm$^{-1}$), which is shifted by only
$-$1.7~km~s$^{-1}$ from the vacuum wavelength of $Q$(3,~0). However,
the feature is not detected in the 2003 May data in a spectrum with
two times higher signal-to-noise ratio. The UKIRT spectrum at the
$Q$(3,~0) line from 2005 March (Fig. \ref{f6}) could be interpreted
as containing a weak and broad emission whose center is near the
expected wavelength of the line. However, no narrow line with strength
approaching that reported by BR02 is present.

\begin{figure*}[tbh]
\begin{center}
\includegraphics[angle=0,scale=0.45]{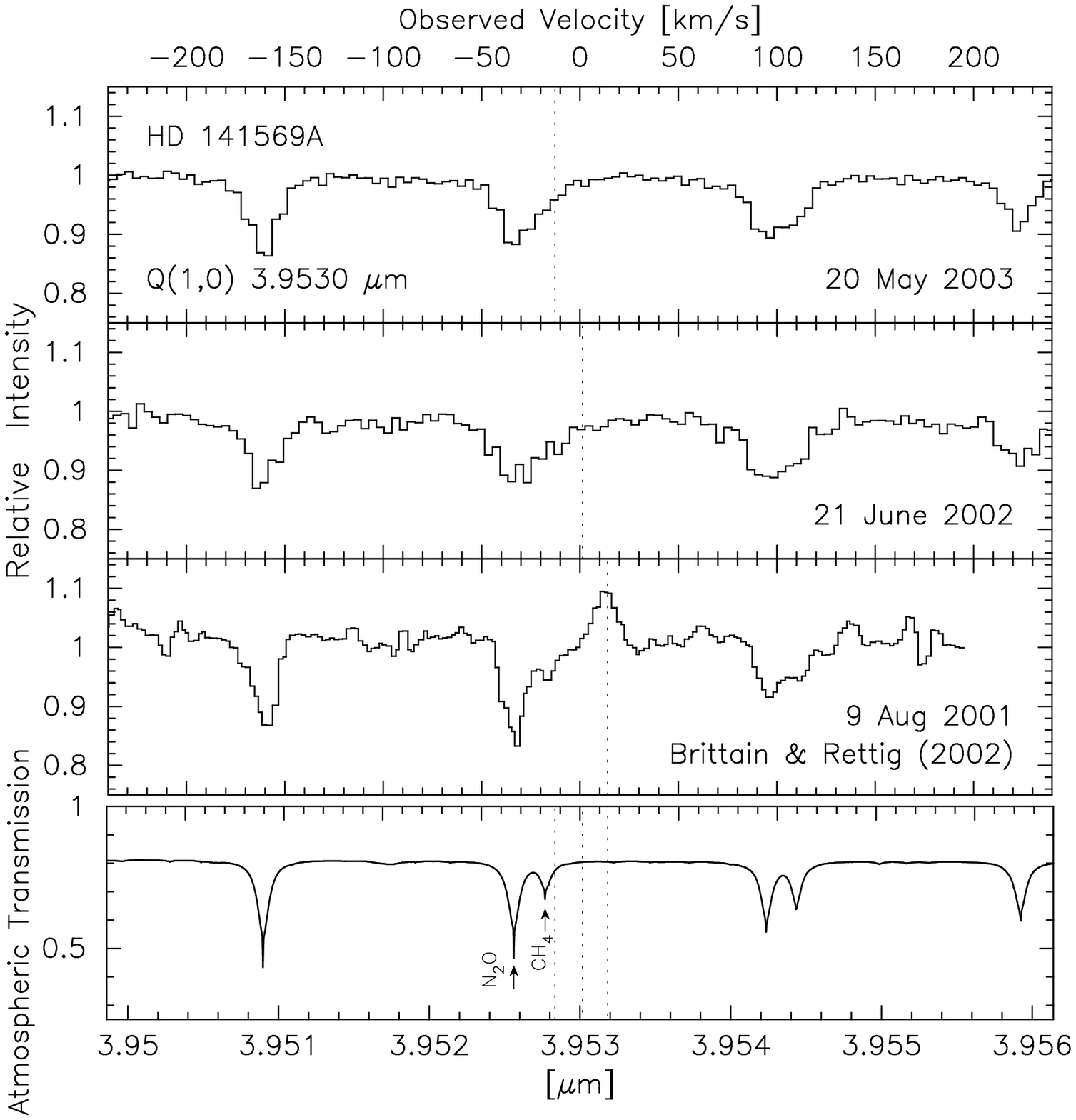}
\hspace{1.0cm}
\includegraphics[angle=0,scale=0.45]{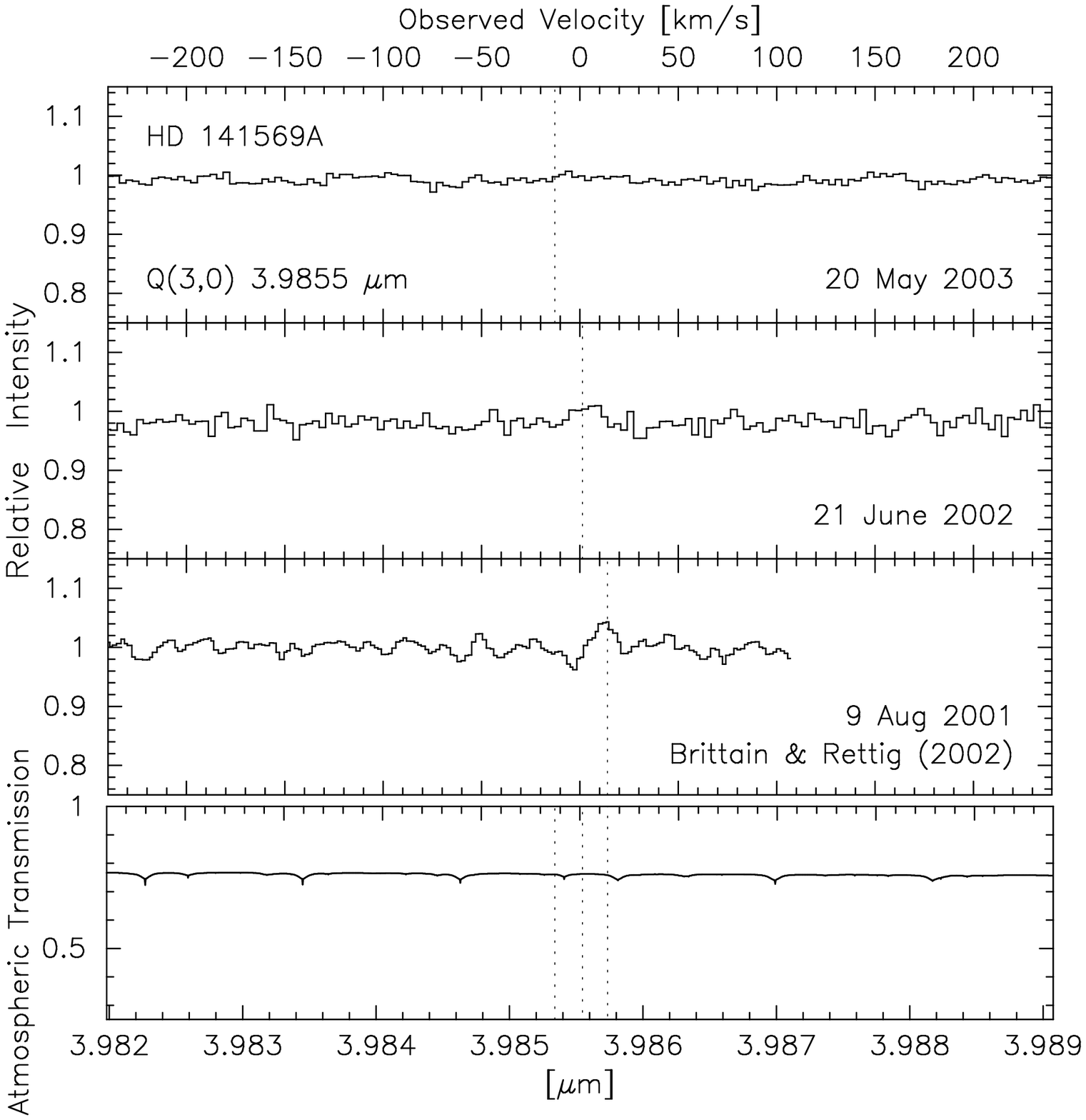}
\caption{Detailed raw spectra of HD~141569A at the wavelengths of
the H$_3^+$ $Q$(1,~0) and $Q$(3,~0) together with the spectra of
BR02. The north-south and east-west Subaru spectra obtained in June 2002
have been summed. Dotted vertical lines denote the wavelengths that the
$Q$(1,~0) and $Q$(3,~0) would appear in May 2003 and June 2002.
The bottom panels show the atmospheric transmission curves computed using
ATRAN (Lord 1992).  The identifications of the absorbing molecules were
made with HITRAN database (Rothman et al. 2003). 
\label{f7}}
\end{center}
\end{figure*}

The Subaru spectra at the $Q$(1,~0) and $Q$(3,~0) lines are more
closely compared with those of BR02 in Fig. \ref{f7} in order to
examine the possible role of interference from telluric lines. In this
figure the raw Subaru spectra are presented, prior to correction for
atmospheric absorption, in order to make a direct comparison with BR02
Fig.~2, in which telluric absorption features are not removed. On 2001
August 9, the line identified as $Q$(1,~0) by BR02 was found at
3.95318~$\mu$m, redshifted by 47~km~s$^{-1}$ from a strong absorption
line of N$_2$O (3.95256~$\mu$m), and by 31~km~s$^{-1}$ from a weaker
line of CH$_4$ (3.95277~$\mu$m). The redshift of this line from the
N$_2$O and CH$_4$ lines would have been 18~km~s$^{-1}$ and
3~km~s$^{-1}$, respectively, on 2003 May 20. It is possible that
H$_3^+$ $Q$(1,~0) emission is partially blocked by these lines, but
neither of the lines is opaque. The overlap would have been less
severe on 2002 June 21, with shifts of 34~km~s$^{-1}$ to N$_2$O and
18~km~s$^{-1}$ to CH$_4$. Although the intensity of the emission line
could be weaker than that observed by BR02, we believe $Q$(1,~0)
should have been detectable at least at the shoulder of CH$_4$ with
the present spectral resolution. Overlapping telluric absorption lines
do not explain the absence of $Q$(3,~0) which is in a clean region,
let alone other transitions, most of which are only modestly affected
by atmospheric absorption lines.

One must also consider the possibility that the H$_3^+$ line emission
could be extended, or localized at an off-center location of the star
and detected only in the wider (1\arcsec) slit used by BR02. However,
both the UKIRT 2001 September and Subaru June 2002 observations
utilized the natural seeing and thus sampled the circumstellar
material more or less evenly, without detecting line emission. The May
2003 Subaru observation was assisted by adaptive optics which achieved
near-diffraction-limited image quality (point spread function 0\farcs1
FWHM at 3.9~$\mu$m). Contamination of the central star spectrum by
detached H$_3^+$ emission would have been minimal. The spectra
obtained from wide aperture (1\arcsec) centered on the star show no
significant difference except somewhat increased noise levels
(Fig. \ref{f3}). From that spectrum we can only rule out significant
H$_3^+$ line emission due north or south of the central star in
2003. However, at that time we also undertook 5-position slit-scanning
spectroscopy within 0\farcs38 of the central star, scanning from east
to west (Fig. \ref{f1}). The data were combined except at the position
of the central star (Fig. \ref{f4}). If asymmetric off-center line
emission from H$_3^+$ contaminates the central star by as much as 10\%
of the continuum level as reported for the $Q$(1,~0) line by BR02, the
offset emitting spot should be very bright and readily detected.
However, no line emission was seen and the spectrum is dominated by a
residual flat continuum from the central star. Thus asymmetrically
distributed H$_3^+$ line emission cannot be the cause of the
discrepant results.

To summarize, we did not detect either of the two lines originally
reported by BR02, at levels significantly lower than they
reported. Rettig (private communication) continues to detect a line at
the wavelength of the $Q$(3,~0) line of H$_3^+$ with the Keck
telescope at close to the originally reported strength. The
discrepancy with our data is currently unresolved.

\section{Discussion}

The most stringent 3$\sigma$ upper limits to the fluxes of H$_3^+$
emission lines searched for here, $\sim$~3~$\times$~10$^{-19}$~W~m$^{-2}$,
correspond to upper limits of $\sim$~3~$\times$~10$^{19}$~W to
luminosities in these lines. These limits are more than eight orders of
magnitude greater than those from Jupiter's brightest lines. Thus H$_3^+$
line emission from a Jupiter-like exoplanet in a Jupiter-like orbit is not
likely to be detectable in the near future.

\figurenum{7}

Brittain et al. (2003) detected emission lines near 5~$\mu$m from CO
in HD~141569A. The emitting CO is in a range of vibrational states
which they argued are excited by UV photons at the inner edge of the
17~AU disk. The H$_3^+$ lines reported by BR02 are approximately an
order of magnitude weaker than the strongest CO lines. In dense
clouds, where H$_3^+$ is created following collisional ionization of
H$_{2}$ by cosmic rays,
$n$(H$_3^+$)~$\sim$~1~$\times$10$^{-4}$~cm$^{-3}$ (Geballe 2000) is
many orders of magnitude less than $n$(CO). One would expect line
emission from vibrationally excited H$_3^+$ to be weaker than that
from vibrationally excited CO, roughly by the ratio of their
abundances, and therefore the H$_3^+$ lines to be undetectable.  In
principle ionization of H$_2$ by UV photons could augment the
abundance of H$_3^+$. However, that requires photons of energy at
least 15.3~eV, higher than those that ionize atomic hydrogen. Such
photons would be shielded from all but the extreme inner edge of the
disk. Thus, whatever process produces the lines of excited CO from the
circumstellar disk should not produce detectable H$_3^+$ line
emission.

BR02 hypothesized that their reported H$_3^+$ line emission could
arise in a gas giant protoplanet.  We consider the conditions required
for such emission, at the upper limit we observed, which corresponds
to 6~$\times$~10$^{38}$ photons~s$^{-1}$ in a single line. We assume,
as did BR02, that the emission is from a gaseous sphere of mass five
times that of Jupiter that is 7 AU from HD~141569A. We assume a
diameter of 1~AU for such an object (60\%\ of the diameter of a Hill
sphere, assuming a $3M_\odot$ star, appropriate for the spectral type
of HD~141569A). The mean number density of such an object is
$\sim$10$^{15}$~cm$^{-3}$.  Assuming that interstellar conditions
pertain for the creation and destruction of H$_3^+$,
($n$(H$_3^+$)$\sim$10$^{-4}$~cm$^{-3}$; Geballe 2000), the sphere
contains $\sim$2~$\times$~10$^{35}$ H$_3^+$ ions. The vibrational
excitation mechanism for the H$_3^+$ is unknown, but as discussed
above cannot be absorption by UV photons. Here we consider the
possibility of collisional excitation. If, for example, the
temperature of the sphere is 600~K, the H$_3^+$ would be vibrationally
pumped by collisions at a rate of $\sim$4~$\times$~10$^{3}$~s$^{-1}$
(the rate coefficient is $\sim$2~$\times$~10$^{-9}$~cm$^{3}$~s$^{-1}$
and the efficiency of vibrational pumping is
$e^{-E/kT}$~$\sim$~2~$\times$~10$^{-3}$), and would emit with an
efficiency of $\sim$6~$\times$~10$^{-5}$ (the ratio of spontaneous
emission to collisional relaxation), so that the photon emission rate,
spread over a few tens of rovibrational lines, is
$\sim$1~$\times$~10$^{35}$~s$^{-1}$. This is roughly five orders of
magnitude below our observed upper limit. Thus, even at this probably
unrealistically high temperature one would need to invoke an H$_3^+$
production rate $\sim$10$^{5}$ times higher than the interstellar
value, or alternatively severe depletions of {\it all} atomic and
molecular species with which H$_3^+$ can easily react in order to
reduce the overall reaction rate by $\sim$10$^{5}$. At lower
temperatures the pumping rate by collisions would drop steeply and an
alternate excitation mechanism would be required.

Is the observational limit derived here a useful constraint on jovian
exoplanets in much closer proximity to their central stars than the
putative protoplanet of BR02 is from HD~141569A, or Jupiter is from the
sun? Miller et al. (2000) estimated that a gas giant planet of jovian mass
at a distance of 0.05~AU, such as the $\tau$~Boo exoplanet would have
three orders of magnitude higher column density of H$_3^+$ in its
ionosphere due to radiative ionization of H$_{2}$ followed by the rapid
reaction of H$_2^+$ and H$_{2}$. One must also add ionization by the
vastly denser stellar wind (10$^{4}$ times denser than at the distance of
Jupiter). Countering the increased sources of H$_3^+$ will be sinks such
as dissociative recombination of H$_3^+$ on electrons. Moreover, in the
extreme environment in which the exoplanet's atmosphere is located, it is
likely that a large fraction of the hydrogen will be atomic and thus
unable to readily form H$_3^+$.

Miller et al. (2000) estimated that H$_3^+$ line fluxes emitted from the
$\tau$~Boo ($d$~=~19~pc) exoplanet could be a
few~$\times$~10$^{-21}$~W~m$^{-2}$, a value two orders of magnitude
lower than our upper limit, but nominally less than an order of
magnitude lower when the difference in distance is included. However,
their estimate did not take into account the effects mentioned above
that would significantly reduce the H$_3^+$ abundance.  Moreover the 25
times brighter continuum flux density of $\tau$~Boo than HD~141569A
makes detection of a line of even this strength considerably more
difficult there. Thus, we expect that the limit on H$_3^+$ luminosity
reported here is several orders of magnitude above meaningful upper
limits for even the most extreme gas giant exoplanet environments. It is
possible that a giant protoplanet, perhaps somewhat more distant that
0.05~AU such that a significant fraction of its hydrogen is molecular,
could glow more brightly in H$_3^+$ than a close exoplanet, but it is
unlikely that any such object is currently in the solar neighborhood. In
summary, we believe that detection of H$_3^+$ line emission from either
type of object remains a formidable technical challenge.

\section{Conclusion}

We have searched unsuccessfully for line emission from H$_3^+$ in
HD~141569A, obtaining upper limits significantly below the signal
strengths reported by BR02, both for the lines that they reported to have
detected and for additional H$_3^+$ lines which one would expect would be
of comparable strength.  We also do not see any other lines in our data,
which cover one-third of the 3.2-4.0~$\mu$m interval. Apart from a
combination of time-variability and a highly unusual and unknown
excitation mechanism, we can find no viable explanation for the conflict
between our results and those of BR02. We have found no plausible
mechanism to produce, in either a gas giant protoplanet or in an
exoplanet, H$_3^+$ luminosities approaching those reported by BR02 or our
upper limits.

\acknowledgments

We thank all the staff and crew of the Subaru Telescope and NAOJ for
their valuable assistance in obtaining these data and continuous
support for the construction of IRCS and Subaru AO system. We also
thank the staff of the Joint Astronomy Centre for its support of the
observations at UKIRT.  Special thanks goes to H. Izumiura and
M. Yoshida at Okayama Observatory for their indispensable help for
obtaining supplementary data on HD~141569. B. J.  M. has been
supported by the Miller Institute for Basic Research in Science.
T. O. is supported by NSF grant PHY-0354200. T. R. G.'s research is
supported by the Gemini Observatory, which is operated by the
Association of Universities for Research in Astronomy, Inc., on behalf
of the international Gemini partnership of Argentina, Australia,
Brazil, Canada, Chile, the United Kingdom and the United States of
America. M. G. is supported by a Japan Society for the Promotion of
Science fellowship.


\end{document}